\documentclass[conference]{IEEEtran}

\ifCLASSINFOpdf
\else
\fi

\usepackage{amssymb}
\def\mymedskip{\vskip\medskipamount}
\def\mymedbreak{\par \ifdim\lastskip<\medskipamount
  \removelastskip \penalty-100 \mymedskip \fi}
\def\myaftermedspace{\par \ifdim\lastskip<\medskipamount
  \removelastskip \penalty55\mymedskip\fi}
\newcommand{\eop}{{\unskip\nobreak\hfil\penalty50
          \hskip2em\hbox{}\nobreak\hfil$\Box$
          \parfillskip=0pt \finalhyphendemerits=0 \par}}
\newenvironment{proof}%
{\mymedbreak{\noindent\bf Proof:\enspace}}{\eop\myaftermedspace}
{\mymedbreak{\noindent\bf Proof of Theorem #1:\enspace}}{\eop\myaftermedspace}
{\mymedbreak\noindent{\bf Remark:}%
\enspace\rm}%
{\myaftermedspace}
\newtheorem{teor}{Theorem}[section]
\newtheorem{defi}[teor]{Definition}
\newtheorem{fact}[teor]{Fact}
\newtheorem{problem}{Problem}
\newtheorem{exercise}{Exercise}
\newtheorem{examp}[teor]{Example}
\newtheorem{lem}[teor]{Lemma}
\newtheorem{cor}[teor]{Corollary}
\newtheorem{con}[teor]{Conjecture}
\newtheorem{prop}[teor]{Proposition}
\newtheorem{rem}[teor]{Remark}
\newcommand{\beq}{\begin{equation}}
\newcommand{\eeq}{\end{equation}}
\newcommand{\beql}[1]{\begin{equation} \label{#1}}
\newcommand{\eeql}{\end{equation}}
\newcommand{\beqa}{\begin{eqnarray*}}
\newcommand{\eeqa}{\end{eqnarray*}}
\newcommand{\beqal}[1]{\begin{eqnarray} \label{#1}}
\newcommand{\eeqal}{\end{eqnarray}}
\newcommand{\beqan}{\begin{eqnarray}}
\newcommand{\eeqan}{\end{eqnarray}}
\newcommand{\bpf}{\begin{proof}}
\newcommand{\epf}{\end{proof}}
\newcommand{\ben}{\begin{enumerate}}
\newcommand{\een}{\end{enumerate}}
\newcommand{\bit}{\begin{itemize}}
\newcommand{\eit}{\end{itemize}}

\newcommand{\cA}{{\cal A}}

\newcommand{\cG}{{\cal G}}

\newcommand{\cU}{{\cal U}}

\newcommand{\gs}{\sigma}

\newcommand{\ga}{\alpha}
\newcommand{\gb}{\beta}
\newcommand{\gc}{\gamma}
\newcommand{\gd}{\delta}
\newcommand{\gre}{\epsilon}

\newcommand{\bab}{\begin{abstract}}
\newcommand{\eab}{\end{abstract}}
\newcommand{\bke}{\begin{keywords}}
\newcommand{\eke}{\end{keywords}}

\newcommand{\btm}[1]{\begin{teor} \label{#1}}
\newcommand{\etm}{\end{teor}}
\newcommand{\btmn}[2]{\begin{teor}[#1] \label{#2}}
\newcommand{\etmn}{\end{teor}}
\newcommand{\ble}[1]{\begin{lem} \label{#1}}
\newcommand{\ele}{\end{lem}}
\newcommand{\bLe}[1]{\begin{Lemma} \label{#1}}
\newcommand{\eLe}{\end{Lemma}}
\newcommand{\bpn}[1]{\begin{prop} \label{#1}}
\newcommand{\epn}{\end{prop}}
\newcommand{\bex}[1]{\begin{examp} \label{#1}}
\newcommand{\eex}{\end{examp}}
\newcommand{\bde}[1]{\begin{defi} \label{#1}}
\newcommand{\ede}{\end{defi}}
\newcommand{\bco}[1]{\begin{cor} \label{#1}}
\newcommand{\eco}{\end{cor}}
\newcommand{\bcorn}[2]{\begin{cor}[#1] \label{#1}}
\newcommand{\ecorn}{\end{cor}}
\newcommand{\bcon}[1]{\begin{con} \label{#1}}
\newcommand{\econ}{\end{con}}
\newcommand{\bfa}[1]{\begin{fact} \label{#1}}
\newcommand{\efa}{\end{fact}}
\newcommand{\bpr}[1]{\begin{problem} \label{#1}}
\newcommand{\epr}{\end{problem}}
\newcommand{\bexer}[1]{\begin{exercise} \label{#1}}
\newcommand{\eexer}{\end{exercise}}
\newcommand{\bre}[1]{\begin{rem} \label{#1}}
\newcommand{\ere}{\end{rem}}

\newcommand{\bbF}{\mathbb{F}}

\newcommand{\bbZ}{\mathbb{Z}}

\newcounter{question_number}

\newenvironment{question}{\addtocounter{question_number}{1}\noindent{\bf Question \arabic{question_number}}}{\myaftermedspace}

\newenvironment{solution}{\noindent {\bf Solution:} \enspace}{\eop\myaftermedspace}

\newenvironment{hint}{\noindent {\bf Hint:} \enspace}{\eop\myaftermedspace}

\newenvironment{multisolution}[1]{\noindent {\bf Solution #1:} \enspace}{\eop\myaftermedspace}

\newcommand{\bqu}{\begin{question}}
\newcommand{\equ}{\end{question}}
\newcommand{\bs}{\begin{solution}}
\newcommand{\es}{\end{solution}}
\newcommand{\bh}{\begin{hint}}
\newcommand{\eh}{\end{hint}}
\newcommand{\bms}[1]{\begin{multisolution}{#1}}
\newcommand{\ems}{\end{multisolution}}

\begin{document}
%
\title{Characterizations and construction methods for linear functional-repair storage codes}
%

\author{
  \IEEEauthorblockN{ Henk D.L. Hollmann and Wencin Poh}
  \IEEEauthorblockA{School of Physical and Mathematical Sciences\\
    Nanyang Technological University\\
    Singapore\\
    Email: \{lhenk.hollmann, wcpoh\}@ntu.edu.sg} 
}

\maketitle
\begin{abstract}
We present a precise characterization of linear functional-repair storage codes in terms of {\em admissible states\/}, with each state made up from a collection of vector spaces over some fixed finite field. To illustrate the usefulness of our characterization, we provide several applications. We first describe a simple construction of functional-repair storage codes for a family of code parameters meeting the cutset bound outside the MBR and MSR points; these codes are conjectured to have optimal rate with respect to their repair locality. Then, we employ our characterization to develop a construction method to obtain  functional repair codes for given parameters using symmetry groups, which can be used both to find new codes and to improve known ones. As an example of the latter use, we describe a beautiful functional-repair storage code that was found by this method, with parameters belonging to the family investigated earlier, which can be specified in terms of only eight different vector spaces.

\end{abstract}


%
\IEEEpeerreviewmaketitle

\section{Introduction}
A distributed storage system (DSS) typically stores data objects in encoded form on multiple storage units, commonly referred to as {\em storage nodes\/}. 
Over time, the DSS will have to handle the occasional loss of storage nodes, for example due to hardware or software failures or peer churning (in peer-to-peer storage systems). This is usually referred to as the {\em repair problem\/}.
Under the simplest repair regime, each data block on a failed node has to be reconstructed {\em exactly\/} and stored on a newcomer node.  Greater efficiency (e.g., higher rates) can sometimes be achieved by employing a repair regime called {functional repair\/}, where the replacement blocks do not need to be an exact copy of the lost data blocks, but are merely required to contain sufficient information to maintain data integrity (in a moment, this will be discussed in more detail).
A  {\em storage code\/} is the precise description of how the DSS handles the data storage and repair management. 

Many different performance criteria for storage code efficiency have been considered. The {\em repair bandwidth\/}, the total amount of data traffic needed during repair, has received the most attention until now, and is currently the best understood. Another performance measure is the repair locality, the number of nodes that need to be contacted during repair.  In applications such as cloud storage or deep archival, minimizing disk I/O seems to be the main consideration. Since the disk I/O is proportional to the number of nodes contacted during repair, the repair locality has emerged as an important parameter and has been much investigated recently. 

Under the functional repair regime, lost data blocks have to be replaced by ``functionally equivalent'' blocks, in the following sense. The obvious requirement is that after repair the original information that was stored can still be retrieved. A less obvious demand is that the replacement must  guarantee that also {\em future\/} (functional) repairs remain possible. Typically, the new data block is replaced by a linear combination of some of the data stored on a subset of the other blocks, in such a way that ``enough independency'' is maintained in the system. There are several problems with this description. First of all, it seems that some kind of ``invariant'' needs to be maintained in the system, but it is not clear what this invariant should be. Secondly, in contrast with an exact repair regime where for a given stored data file only a limited number of different data blocks can occur  and each node has a clear functional identity,  under a functional repair regime, in principle every possible data block might appear and nodes do not have a clear identity. These facts complicate the system management, and make code design less evident, especially for small repair locality. Indeed, few codes for the functional repair regime are known \cite{StorageWiki}. 

In this paper, we present a rather general method to construct functional-repair storage codes for given parameters and symbol alphabet size. To this end, we first offer a description of (exact and functional repair) {\em  linear\/} storage codes in terms of vector spaces over finite fields. We remark that most known codes can be described in this way. 
One of the advantages offered by  such a description is that it emphasizes the  underlying structure of the storage code. This opens the way to investigate and construct storage codes with a certain common structure, similar to the situation in the theory of error-correcting codes, where one studies for example  cyclic codes  or other highly structured codes, thus facilitating code design or correction management. Here, we use this description to clarify the idea of a hidden ``invariant'' that has to be maintained, in a way that enables, in principle, to decide on the existence or non-existence of a linear storage code for given ``small'' parameters.  The idea here is to characterize the code in terms of {\em admissible  states\/}, where each state  is a collection  of vector spaces that essentially describes a possible set of data blocks stored on the nodes at some moment in time.
Finally, we describe a method to use symmetry groups to search for small storage codes (having a small set of admissible states) with a high degree of symmetry (technically, having a transitive automorphism group). Search methods for exact-repair storage codes are described for example  in~\cite{CDHmin}, but as far as we know search methods for functional-repair storage codes have not been published before. 

To illustrate our ideas, we investigate a class of storage code parameters that has been proved or conjectured to provide optimal rate for a given repair locality \cite{Holl1301:Storage}. In most cases, storage codes with these parameters are necessarily functional-repair. Since these codes can be considered as regenerating codes, it is known that functional repair codes with these parameters exist provided that the symbol alphabet size is large enough. We use our methods to give a simple, explicit construction for these codes, which enables the use of our group-theoretical construction techniques to search for smaller codes with the same parameters. As an example, we describe one such code thus found that can be described in terms of only eight different binary vector spaces.
 
For an overview of DSS and storage codes, we refer to \cite{nc-survey} or \cite{overview-ddsc}, or to the Storage Wiki \cite{StorageWiki}.



\section{Storage codes}
A {\em storage code\/} with parameters $(m,n,k, r, \ga, \gb)$ is a code that allows {\em resilient\/} storage of $m$ information symbols from some finite alphabet $\bbF$, in encoded form, onto $n$ storage nodes, each capable of holding $\ga$ data symbols from $\bbF$. We will refer to $\ga$ as the {\em storage node capacity\/}. The parameter $k$ indicates that at all times, the original stored information can be recovered from the data stored on {\em some\/} set of $k$ nodes. If {\em any\/} set of $k$ nodes will do, the the code is referred to as MDS \cite{mw-sl}. The {\em rate\/} of the code is  the fraction $m/(n\ga)$ of information per stored symbol. 
The resilience of the code is described in terms of a parameter $r$, referred to as the {\em repair locality\/}, and a parameter $\gb$, referred to as as the {\em transport capacity\/} of the code. If a node fails, then a newcomer node is allowed to contact {\em some\/} set of $r$ live nodes, called the {\em repair set\/}; each of these $r$ nodes computes an amount of $\gb$ data symbols, which are then downloaded by the newcomer node  in order to regenerate some of the lost information, in the form of a replacement block again consisting of $\ga$ data symbols. If the code is {\em exact-repair\/}, then we require that this replacement block is an {\em exact\/} copy of the lost data block. However, for certain purposes this repair modality is too strict, and it is convenient to consider a weaker form of resilience.  We say that the code allows {\em functional repair\/} if the replacement block is {\em information equivalent\/} to the lost data block, while ensuring the possibility of future functional repair of other nodes. The notion of functional repair is more subtle and much more difficult to grasp; we will provide more explanation and various examples later in this paper. 
The storage code describes the entire data and repair management of the DSS.


%
In \cite{Holl1301:Storage}, a framework has been developed for the description of linear storage codes in terms of vector spaces over a finite field. Given a linear storage code for the exact repair regime, it is 
rather straightforward to construct  such a description, however for linear storage codes under functional  repair, things are less clear. In the
remainder of this section, we briefly review the relevant notions from~\cite{Holl1301:Storage}; in the next section, we will use these ideas to derive a characterization of linear functional repair storage codes.
%

\subsection{Linear storage codes}
In this paper, we will think of a {\em linear\/} exact-repair storage code as a collection of~$n$ subspaces $U_1, \ldots, U_n$ of an $m$-dimensional vector space $\bbF^m$, each of dimension~$\ga$. We will refer to $\bbF^m$ as the {\em message space\/} and to the $U_i$ as the {\em storage node spaces\/} or, more briefly, as the {\em node spaces\/}. The integer $\ga$ is called the {\em (storage) node capacity\/}.

A {\em recovery set\/} of the storage code is a subset of the storage node spaces that together span the entire message space~$\bbF^m$.
Here, the {\em span\/} of a collection of vector spaces $W_1, \ldots, W_k$  is the collection of all vectors $w_1+\cdots w_k$ with $w_i\in W_{i}$ for all $i$, that is, the smallest vector space containing all the vector spaces $W_1, \ldots, W_k$.
The {\em  recovery dimension\/} of the storage code is defined as the {\em smallest\/} size~$k$ of the a recovery set.

A linear storage code as above can be used to store an amount of $m$ data symbols in $n$ storage nodes, in the following way. Associate the~$n$ storage nodes $v_1, \ldots, v_n$ with the~$n$ storage spaces $U_1, \ldots, U_n$, and choose a fixed basis in each of them. Now, represent the data to be stored as an vector $x\in\bbF^m$; then, in node $v_i$, we store the $\ga$ inner products 
of $x$ with the $\ga$ basis vectors 
of~$U_i$. In other words, if $B_i$ is the $m\times \ga$ matrix that has as its columns the basis vectors for~$U_i$, then storage node $v_i$ stores $x^\top B_i$.
Note that, consequently, the DSS can compute the inner product of the data vector~$x$ with any vector contained in one of the node spaces; moreover, from the data stored in a recovery set, which by definition is a spanning subset of the node spaces, the DSS can recover~$x$. 

Now let us consider the repair problem. 
Fix some positive integer~$\gb$, referred to as the {\em transport capacity\/} of the storage code. We will say that a collection $R$ of the   node spaces  is a {\em repair set \/} for a certain  node space $U_\ell\notin R$  if it is possible to choose in each  $U_i\in R$ a $\gb$-dimensional {\em repair space\/} $W_{i,\ell}$  
such that $U_\ell$ is contained in the span of the repair spaces $W_{i,\ell}$. Note that if we represent each repair space $W_{i,\ell}$  by a fixed $m\times \gb$ matrix $B_{i,\ell}$ having the vectors of a fixed basis for~$W_{i,\ell}$ as its columns,  then the vector $x^\top B_\ell$ stored in node $v_\ell$ can be recovered from a linear combination of the vectors $x^\top B_{i,\ell}$, which in turn can be computed from the vectors $x^\top B_i$ stored in the nodes $v_i$ involved in  the repair set. 

If {\em each\/} node space in the code has a repair set  of size~$r$ with respect to  transport capacity $\gb$ then we say that the storage code has {\em repair locality\/} $r$ with respect to transport capacity $\gb$. We will refer to a storage code with all the above parameters as a  {\em linear exact-repair $(m; n,k,r, \ga,\gb)$-storage code\/}. Note that such a storage code will have a coding rate $R=m/(n\ga)$. 
The following simple example illustrates the above notions.
\bex{E42} \rm
Consider the linear storage code with node spaces $U_0=\langle e_0, e_2+e_3\rangle$, $U_1=\langle e_1, e_3+e_0\rangle$, $U_2=\langle e_2, e_0+e_1\rangle$, and $U_3=\langle e_3, e_1+e_2\rangle$, considered as subspaces of $\bbF_2^4$. Here, we write $\langle w_1, \ldots, w_k\rangle$ to denote the span of the vectors $w_1,\ldots, w_k$, the vector space consisting of all linear combinations of $w_1,\ldots, w_k$. We claim that these node spaces constitute an $(m=4; n=4,k=2,r=3,\ga=2,\gb=1)$ linear exact-repair storage code. Indeed, there are $n=4$ node spaces $U_i$, each of dimension $\ga=2$. 
Furthermore, $k=2$ since 
any two subspaces intersect trivially, so together span the entire space $\bbF_2^4$.
Furthermore, the set $R=\{U_1, U_2, U_3\}$ is a repair set for node space~$U_0$, with respect to transport capacity~$\gb=1$. Indeed, if we choose repair spaces $W_{1,0}=\langle e_0+e_3\rangle \subseteq U_1$, $W_{2,0}=\langle e_2\rangle \subseteq U_2$, and $W_{3,0}=\langle e_3\rangle \subseteq U_3$ (each of dimension $\gb=1$), then $U_0\subseteq \langle e_0+e_3,e_2, e_3\rangle=W_{1,0}+W_{2,0}+W_{3,0}$ as required.
The storage code has rotational symmetry: the linear transformation given by $e_i\mapsto e_{i+1}$ (indices modulo 4) maps $U_i$ to $U_{i+1}$; as a consequence, repair sets for the other node spaces can be obtained by symmetry.
With the bases as suggested by the above description, this code stores a data vector $x=(x_0, \ldots, x_3)$ by letting node 0 hold $x_0$ and $x_2+x_3$,
(and letting node $i$ hold $x_i$ and $x_{i+2}+x_{i-1}$ for $i=1,2,3$),  
and repairs node 0 by downloading $x_0+x_3$ from node 1, $x_2$ from node 2, and $x_3$ from node 3. (Storage and repair for the other nodes follows by symmetry.)
\eop
\eex

\subsection{Linear functional-repair storage codes}
The linear exact-repair storage codes in the previous section had the property that a {\em fixed\/} vector space was associated with each node. For linear {\em functional-repair\/} storage codes, this will no longer be the case.
Recall that under the regime of functional repair, a data block on a failed storage node has to be replaced by a data block on a newcomer that is {\em information equivalent\/} to the one on the failed node, while ensuring the possibility of future functional repair of other nodes. Linear functional-repair storage codes  are perhaps best thought of as a specification of {\em admissible\/} node space arrangements, with the property that in every such arrangement, a node space can be ``repaired'' by replacing it with a (possibly different) space so that the resulting arrangement again satisfies the specifications. Consider the following example.
\bex{MBR-func}\rm
We will construct a linear functional-repair storage code with parameters $(m=5; n=4, k=r=3, \ga=2, \gb=1)$,  so with coding rate $R=5/8$.
As message space we take~$\bbF_2^5$. We will ensure that at any moment, the four 2-dimensional  node spaces $U_1, \ldots, U_4$ associated with the four storage nodes satisfy the following specification:
\begin{enumerate}
\item Any two of the node spaces intersect trivially, that is, $U_i\cap U_j=\{0\}$ when $i\neq j$;
\item Any three of the node spaces span the entire message space $\bbF^5$. 
\end{enumerate}
Assuming that $U_1, \ldots, U_4$ satisfy these constraints, suppose that node 4 fails. 
Without loss of generality, we may assume that  
$U_1=\langle e_1, a_1\rangle$, $U_2=\langle e_2, a_2\rangle$, and $U_3 = \langle e_3, a_3\rangle$,  for some basis $e_1, e_2, e_3, a_1, a_2$ of~$\bbF^5$, with $a_1+a_2+a_3=0$. Indeed, $U_3$ must have trivial intersection with both $U_1$ and $U_2$, but has to intersects the 4-dimensional span $U_1+ U_2$, hence this intersection is of the form $a_1+a_2$ with $a_i\in U_i^*=U_i\setminus \{0\}$.  This shows that $U_1, U_2, U_3$ must have the indicated form. Now, to repair (or initially construct) the storage space $U_4$, given that $\gb=1$ we must choose a vector $w_i\in U_i^*$ for $i=1,2,3$, and let $U_4$ be some 2-dimensional subspace of their span $\langle w_1, w_2, w_3\rangle$, which by rule 1 should  not contain any of the $w_i$.  Hence $U_4$ is of the form $\{0, w_1+w_2, w_1+w_3, w_2+w_3\}$. Furthermore, 
$w_3\neq a_1+a_2$ since otherwise $U_4\subset U_1+U_2$, violating rule 2, and similarly, $w_1\neq a_1, w_2\neq a_2$. So $w_1=e_1+x_1 a_1, w_2=e_2+x_2a_2, w_3=e_3+x_3a_3$, and it is now easily verified that any choice of $x_1, x_2, x_3\in \bbF_2$ is valid. 
(Initially, we can take, for example, $U_4=\langle e_1+e_2, e_1+e_3\rangle$.) This shows that we can maintain the specification forever, provided that no two nodes ever fail simultaneously.
\eop
\eex
Using a functional-repair storage codes as above to actually store information is similar to using the exact-repair storage codes introduced earlier, except that now at each moment the other nodes and the data collector have to be informed of the actual {\em state\/} of a storage node, that is, of its current  storage node space, and have to be aware of all the bases used.  This extra overhead can be relatively small if the code is used  to store a large number of messages {\em simultaneously\/}.

\section{Admissible states}
We will now offer a new characterization of linear functional-repair codes in terms of {\em admissible states\/}. Consider again the code in Example~\ref{MBR-func}. The crucial facts making the construction work are 
that the collection $\cU=\{U_1, U_2, U_3\}$ satisfies the specification and that a fourth node space $U_4$ can be constructed  such that any triple from $\cU\cup  \{U_4\}$ again satisfies the specification. In a sense, these collections of triples form the ``admissible repairing collections'' of the code. This  
observation motivates the following two definitions. 
\bde{LDbrep} Let $\cU$ be a collection of $\ga$-dimensional subspaces of a vector space $\bbF^m$. We say that an $\ga$-dimensional subspace $U$ of~$\bbF^m$ can be obtained from $\cU$ by {\em $(r,\gb)$-repair\/} if it is possible to choose $r$ spaces $U_1, \ldots, U_r$ in~$\cU$, the {\em repair set\/}, and then 
a $\gb$-dimensional subspace $W_i\subseteq U_i$ in each of them, the {\em repair spaces\/}, such that $U$ is contained in the span $W_1+\cdots +W_r$ of the repair spaces. 
\ede
%
\bde{LDfrsc} A linear functional repair storage  code with parameters $(m; n, k, r, \ga, \gb)$  over some finite field~$\bbF$ is a set $\cA$
of $(n-1)$-tuples $\cU$ of $\ga$-dimensional subspaces of~$\bbF^m$, such that the following {\em repair property\/} holds. 
Given any $(n-1)$-tuple $\cU=\{U_1, \ldots, U_{n-1}\}$ in $\cA$, there exists an $\ga$-dimensional space $U$ that can be obtained from~$\cU$  by $(r,\gb)$-repair such that  for every $i=1, \ldots, n-1$, the $(n-1)$-tuple $\cU\cup\{U\}\setminus \{U_i\}$ is again in~$\cA$.
Furthermore, we require that each $(n-1)$-tuple $\cU$ in~$\cA$ contains a {\em spanning subset\/} of size $k$, that is, there can be found $k$~spaces in~$\cU$ that together span~$\bbF^m$.
\ede
We will sometimes refer to the $(n-1)$-tuples $\cU$ in~$\cA$ as the
{\em admissible repairing collections\/} of the code, and to the pairs $(\cU,U)$ as the {\em admissible states\/} of the code. We invite the reader to verify that an exact-repair storage code  with parameters as in Definition~\ref{LDfrsc} and storage node spaces $U_1, \ldots, U_n$ is also of this type, with as admissible states $(\{U_1, \ldots, U_n\}\setminus \{U_i\}, U_i)$ for $i=1, \ldots, n$. Returning to the code in Example~\ref{MBR-func}, we recognize it to also be of this type, with as admissible repairing collections all triples $\cU=\{U_1, U_2, U_3\}$ of 2-dimensional node spaces in~$\bbF^5$ that together span ~$\bbF^5$ in which any pair of node spaces intersecting trivially.

We hope that the following characterization  result is  evident by now. For a full explanation, we refer to \cite{hdlh-wp-full}.
\btm{LTfrsc}
Every linear functional-repair code according to Definition~\ref{LDfrsc} is indeed a storage code under the regime of fuctional repair. Conversely, every linear functional-repair code 
can be obtained  from a collection $\cA$ of admissible repairing collections with the properties as in Definition~\ref{LDfrsc}.
\etm
The above results open the way to construction methods for functional-repair codes, and methods to find better, smaller such codes, as we will illustrate in the remainder of this paper.  To the best of our knowledge, a description of functional-repair codes in terms of  ``admissible states'' is new. 
In a few published  constructions for functional-repair codes, some notion of state can be recognized \cite{shum-hu-isit}, \cite{HLS-func}. 


\section{A family of functional-repair codes} 
In this section, we will illustrate the usefulness of our characterization of linear functional-repair codes developed above by proving the existence of a family of codes with parameters $(m_{r,s}; n=r+1, k=r, r, \ga=s+1, \gb=1)$ and rate $(r-s/2)/(r+1)$, where $m=m_{r,s} = (r-s)(s+1) +s +(s-1)+\cdots + 1 = (r-s)(s+1) +{s+1\choose 2}$ and $r>s\geq 0$.
Note that these parameter sets all meet the cutset bound from \cite{Dimakis},  in points different from the MBR and MSR points. (The code in Example~\ref{MBR-func} is the case where $r=3, s=1$.)  Moreover, these parameter sets also meet a bound for the maximal rate of a storage code with repair locality~$r$, storage node capacity~$\ga$, and transport capacity~$\gb$ conjectured in \cite{hdlh-rlover-isit}.

So the construction idea is to come up with a suitable set $\cA$ of admissible repairing collections, and then to show that this set indeed satisfies the requirements in~Definition~\ref{LDfrsc}. We will say that a collection of $(s+1)$-dimensional vector spaces $\cU=\{U_1, \ldots, U_r\}$ in~$\bbF^m$ with $m=m_{r,s}$ is {\em $(r,s)$-good\/} if the span of any $r-s+j$ of these subspaces has dimension $(r-s)(s+1)+s+\cdots +(s+1-j)$, for every $j=0, \ldots, s$. In order to prove that this indeed provides a suitable set of admissible repairing collections, the following result is crucial. 
\btm{LTmds} Suppose that $\cU=\{U_1, \ldots, U_r\}$ is $(r,s)$-good over $\bbF$. Let $w_i\in U_i$ for $i=1, \ldots, r$, and let $U$ be an $\ga$-dimensional subspace of the span~$W$ of $w_1, \ldots, w_r$. 
Let $C\subseteq \bbF^r$ be the collection of all vectors $c=(c_1, \ldots, c_r)$ for which $\sum_{j=1}^r c_j w_j \in U$. 
Then the collections $\cU\cup\{U\}\setminus \{U_i\}$ for $i=1, \ldots, r$ are all $(r,s)$-good if and only if the vectors $w_1, \ldots, w_r$ are independent and the code $C$ is an $[r,s+1,r-s]$ linear MDS code  over $\bbF$.  
 \etm
For a proof of this result, we refer to \cite{hdlh-wp-full}. Note that the mentioned MDS codes certainly exist for a field size $|\bbF|\geq r-1$, see, e.g., \cite{mw-sl}. It follows immediately from the $(r,s)$-good property that independent vectors $w_1, \ldots, w_r$ as in the lemma can always be found, moreover, they can be used to recursively construct {\em at least one\/} $(r,s)$-good collection for all $r$ and  $s$ with $r> s\geq 0$. Therefore, Theorem~\ref{LTmds} in combination with Theorem~\ref{LTfrsc} proves the existence of linear functional-repair codes for the above parameter sets, with field size equal to the smallest prime power $q$ for which $q\geq r-1$.

\section{\label{LSgroup}A construction method using groups}
One of the drawbacks of functional-repair storage codes is that the set of admissible repairing collections that define the code and the number of admissible states  can be huge, which severely complicates the data management in the DSS. Therefore, it is desirable to find {\em small\/} codes. For example, if possible we would like to find a small {\em subset\/} of admissible repairing collections that itself again defines a code, or we would like to find a small set of admissible states for certain given parameters {\em directly\/}.  We will now briefly sketch a method to 
do so, based on symmetry groups. 


Suppose that we try to find a linear functional-repair storage code containing some ``potential'' admissible state $\gs=(\cU,U)$, so where $U$ can be obtained from $\cU$ by $(r,\gb)$-repair, that we guess to belong to the code. The idea is to construct a storage code with a large symmetry group in which $\gs$  is indeed an admissible state. Our method depends on the following.
\btm{LTgc}
Let $\gs=(\cU=\{U_1, \ldots, U_{n-1}\}, U=U_n)$, 
with each of $U_1, \ldots, U_n$ an $\ga$-dimensional subspace of~$\bbF^m$, such that $U_n$ can be obtained by $(r,\gb)$-repair from~$\cU$, and with $\cU$ containing a spanning $k$-subset if $k<n-1$. Suppose we can find for every $i=1, \ldots, n-1$ an invertible linear map $L_i$  mapping $\cU$ to $\cU\cup \{U_n\} \setminus \{U_i\}$. Then with $\cG=\langle L_1, \ldots, L_n\rangle$, the group generated by $L_1, \ldots, L_n$, the set $\cA$ consisting of all images $G(\cU)$ with $G\in \cG$ is a collection of admissible repairing collections for a linear storage code with parameters  $(m; n, k, r, \ga, \gb)$.
\etm 
For the proof of this theorem, we again refer to~\cite{hdlh-wp-full}. We remark that it is sometimes advisable to construct a ``very regular'' potential (admissible) state $\ga=(\cU,U)$, one for which the stabilizer $\Gamma$ of $\gs$ is transitive on~$\cU$; in that case, it is sufficient to find a single invertible map $L$ mapping $\cU$ to $\cU\cup \{U_n\} \setminus \{U_1\}$. In practice, a search can then be set up to see if there is a small group $\cG$ generated by (a transitive subgroup of) $\Gamma$ and one such $L$. 

There are two essentially different situations in which this theorem can be applied. First, if the desired storage code is {\em not\/} yet known to exist, then we must ``guess'' the form of a typical state for the code, after which the hunt for the group can begin. If there already {\em exists\/} a storage code with the desired parameters, then we just take any state of the code, and then use the theorem to search for a {\em small subcode\/} of the code at hand. 
As an example of the second type, in the next section we will describe  a beautiful functional-repair storage code that was originally  found  by computer using this method. 

\section{A combinatorial description of a small functional-repair code}

A {\em vector space partition\/} for a vector space $V$ is a collection of subspaces $V_1, \ldots, V_m$, not necessarily all of the same dimension, such that each nonzero vector in~$V$ is contained in exactly one of the spaces $V_i$. 
It turns out that our code arises from a vector space partition of Beutelspacher type, see, e.g.,  \cite{He-vsp}. 
Consider the finite field $W=\bbF_{8}$, constructed with a primitive element $\ga$ with $\ga^3=\ga+1$. 
We will consider $W$ as a 3-dimensional vector space over~$\bbF_2$. Note that $W$ has a  2-dimensional subspace $U = \{0,\ga,\ga^2, \ga^4\}$ of~$W$ that is invariant under the Frobenius map $x\rightarrow x^2$. We will first construct a vector space partition for $V=W\oplus U$. To that end, for each $\gb\in \bbF_8$,
we define $U_\gb=\{ (\gb u,u) \mid u \in U\}$. It is easily seen that each $U_\gb$ is a 2-dimensional subspace of~$V$, with $U_0=\{0\}\oplus U$; moreover, $U_\gb\cap U_\gc=\{0\}$ when $\gb\neq \gc$.  Also note that $W=W\oplus \{0\}$ intersects each $U_\gb$ only in the zero vector. There are 31 nonzero vectors in~$V$, 7 of which are in $W$ and another 3 in each of the 8 subspaces $U_\gb$. We conclude that $W$ together with the 8 spaces $U_\gb$ forms a vector space partition.

We will now describe a {\em small\/} linear functional-repair storage code with parameters $(m=5; n=4, k=r=3, \ga=2, \gb=1)$.  Earlier, we constructed a storage code for these parameters in Example~\ref{MBR-func}. Here, the admissible states  of our code will be of the form 
$ \gs=(\cU,U) = (\{U_\gb,U_\gc,U_\gd\}, U_\gre)$
for all subsets $\{\gb,\gc,\gd\}$ of size 3 from~$\bbF_8$; for each such subset $\{\gb,\gc,\gd\}$, there will be a {\em unique\/} $\gre$ such that 
$\gs$is a coding state.  It can be shown that, in fact, the unique choice for $\gre$ is to let $\gre^2=\gb\gc+\gb\gd+\gc\gd$. 

The group $G=A\Gamma L(1, \bbF_8)$ (sometimes written as $\Gamma A_1(8)$) consists of all semi-linear transformations $g: x\rightarrow g(x)= ax^{2^i}+b$ for $a\in \bbF_8\setminus\{0\}$, $b\in \bbF_8$, and $i\in \bbZ_3$. It is not difficult to see that these operations indeed form a group: since $x\rightarrow x^2$ acts linear on~$\bbF_8$, if $g: x\rightarrow ax^{2^i}+b$ and $h: x\rightarrow cx^{2^j}+d$, then $h\circ g: x \rightarrow ca^{2^j} x^{2^{ij}} +(cb^{2^j}+d)$ (note that $\ga^{2^3}=\ga=\ga^{2^0}$, so it is proper to consider $i$ modulo 3). We let $G$ act on the vector space $V$ above by associating with each group element $g: x\rightarrow ax^{2^i}+b$ the linear transformation
$L_g: (w,u)\rightarrow (aw^{2^i}+bu^{2^i}, u^{2^i})$
on~$V$. Since $L_hLg=L_{h\circ g}$, this is a well-defined group action. As a result, we have that $L_g: U_\gb \rightarrow U_{g(\gb)}$.
We remark that the subgroup ${\rm AGL}(1,\bbF_8)$ consisting of the {\em linear\/} transformations in $A\Gamma L(1, \bbF_8)$,  acts regular and 3-homogeneous on~$\bbF_8$, see, e.g., \cite{lw-hom}, which provides an alternative way to quickly check that we indeed obtain a storage code. This code is related to structures described in~\cite{klz-112}.
In fact, this code was originally obtained by the group-theoretical methods described in the previous section, employing an initial state involving an (essentially unique) $(r=3, s=1)$-good collection. The group fixing such a state has a transitive subgroup $\langle T\rangle$, and there are eight candidate maps $L$ mapping the state to a suitable successor state, two of which generate a group $\langle T, L\rangle\cong G=A\Gamma L(1, \bbF_8)$. 
Since all other states are images of the initial state by some linear transformation, we know that 
the 8 subspaces $U_\gb$ of $\bbF_2^5$  ($\gb\in\bbF_8$)  have the property that any two are independent (have intersection $\{0\}$) and any three span the entire space $\bbF_2^5$. 
In fact, this property characterizes the storage code. Indeed, it is possible to show that 8 is the maximum number of 2-dimensional vector spaces in~$\bbF_2^5$ that can have this property; moreover, any such collection of 8 spaces is equivalent under some linear transformation to the 8 spaces $U_\gb$ defined above. For further details on precisely how this code was obtained and for proofs, we refer to \cite{hdlh-wp-full}.

\section{Conclusion}
We have developed a characterization of linear functional-repair storage codes over a finite field $\bbF$ in terms of admissible states and admissible repairing collections of the code, collections of vector spaces over $\bbF$ with precisely described properties. We have illustrated the usefulness of this characterization by using  it to construct functional-repair storage codes for a family of parameters conjectured to have maximal possible rate in terms of repair locality, and two other parameters, the storage node capacity $\ga$ and transport capacity $\gb$; these parameter sets also meet the cutset bound, in a point different from the MBR and MSR points. 
Our new characterization  has also provided a general construction method for linear storage codes employing symmetry groups. Finally, we have described a beautiful small functional-repair storage code with a large, transitive automorphism group obtained with our methods.



\section*{Acknowledgment}
The authors would like to thank Lluis Pamies-Juarez 
and Frederique Oggier 
for  proofreading and providing some useful feedback.
The research of Henk D.L. Hollmann and Wencin Poh is
supported by the Singapore National Research Foundation
under Research Grant NRF-CRP2-2007-03.






%

\end{document}